# Estimation of fusion-evaporation cross sections by artificial neural networks

Serkan Akkoyun

*Sivas Cumhuriyet University, Faculty of Science, Department of Physics, 58140 Sivas, Turkey*

**Abstract**

Accurate determination of fusion-evaporation reaction cross section is important in experimental nuclear physics studies. In this study, by using artificial neural network (ANN) method, we have estimated the cross section values for different reactions. The related root mean square errors have been obtained as 18.5 and 110.4mb for training and test data which correspond to 1.8% and 10.5% deviations from the experimental values, respectively.This order of deviations is lower than the cross section value from most common theoretical calculations. The results of this study indicate that ANN method is capable for the estimation of cross section values of fusion-evaporation reactions.

## 1. Introduction

Heavy-ion fusion-evaporation reactions have attracted much attention because of their usefulness for studying the properties and interactions between complex nuclei.The correct knowledge of the fusion-evaporation reaction cross section is quite important especially in experimental nuclear physics studies such as determination of new reaction parameters for a certain reaction and synthesizing a new heavy element. The experimenter has some choice to use the reaction which is suited for the production of interested nuclei. On the earth 255 stable nuclei exist and all other nuclei are created in the laboratories for the experimental nuclear physics studies. The reaction parameters have to be optimized for achieving the highest production rates of the isotope of interest. By this way, one can prepare appropriate beam-target-energy combination in order to produce desired compound nucleus with higher probability. For this purpose, there are several codes and database exist in the literature (Puhlhofer, 1977; Reisdorf, 1981; Gaimard and Schmidt, 1991; Gavron, 1980; Charity, 2008; Dasso and Landowne, 1987; Fernandez-Niello et al., 1989). But the most reliable cross section data comes from the experiments which are performed before. In the absence of experimental values for a certain reaction, the cross section data calculated only by the codes. Whereas, the codes generally give results 5-10 factor different from the experimental cross-section values (Blank et al., 2018).

Artificial neural network (ANN) is a perfect mathematical tool for the estimation of the values which is related to the different fields in science and technology including nuclear physics studies. Even in the case of highly non-linear relationship between dependent and undependent data points, the method properly works. Recently, ANN has been used in many fields in nuclear physics, such as construction of consistent physical formula for the recoil energy distribution in HPGe detertors (Akkoyun and Yildiz, 2013), determination of one and two proton separation energies (Athanassopoulos et al., 2004), developing nuclear mass systematics (Athanassopoulos et al., 2004), determination of ground state energies of the nuclei (Bayram et al., 2014), identification of impact parameters in heavy-ion (David et al., 1995; Baas et al., 1996; Haddad et al., 1997), determination of beta-decay energies (Akkoyun et al., 2014) and estimating nuclear rms charge radius (Akkoyun et al., 2013).

In this study, we have used a powerfool tool ANN (Haykin, 1999) in order to obtain fusion-evaporation cross section data. The experimental data has been barrowed from the literature (Haider and Malik, 1984). By the knwon experimental cross section values of different beam-target combinations in different beam energies, we have predicted fusion-evaporation reaction cross sections. According to the results, the values from the ANN is close to the experimental data. Also the root mean square error values are lower than the most commonly used code of PACE (Gavron, 1980) in comparison with the realistic experimental cross section data. Therefore, we have concluded that the method is convenient for the estimation of the cross section data.

## 2. Artificial Neural Network

Artificial neural network (ANN) is a mathematical tool that mimics the human brain functionality (Hornik et al., 1989). It includes neurons in different layers which are input, hidden and output layers. Due to the layered structure, this type of ANN is called as layered ANN. Input layer neurons receive data from outside and transmit to hidden and then output layer neurons via adaptive synaptic weighted connections. Because of the forward one-way flow of the data, the ANN is named as layered feed-forward ANN. The neurons are the processing units and connected to each other in different layers via adjustable synaptic weights. The main purpose of the method is the determination of the values of weights by using the given sample data. In the hidden layer, the data is activated by using an appropriate activation function which is commonly used sigmoid-like tangent hyperbolic ($tanhx=(e^x-e^{-x})/(e^x-e^{-x})$) function. The numbers of neurons in the input and output layers depend on the variety of the data. Besides, the numbers of hidden layer and its neurons are related to the problem nature, but generally one hidden layer is sufficient for almost all problems. However, the perfect neuron number in this layer is determined which gives the best results after several trying on the problem.

The ANN method is a perfect tool for both linear and non-linear function approximations. It is composed of two main stages. The whole data belonging to the problem is divided into two separate sets for these stages. In the training stage which is the first stage, the first part of data is given to the ANN including both input and desired output values. The weights are modified by using the sample data in the training stage. The method generates its own outputs as close as possible to the desired output values. The comparisons between desired output and ANN output are made by root mean square error (RMSE) function given by Eq.1.

$$RMSE = \sqrt{\frac{\sum_{i=1}^{N}(y_i-f_i)^2}{N}} \tag{1}$$

where $N$ is the total number of the data in the stage, $y_i$ is desired output while $f_i$ is ANN output. After an acceptable agreement between the ANN outputs and the desired outputs, the training stage is finally terminated. It means that the ANN is constructed for solving of the problem with the modified final weights. However, it is still early to decide whether the constructed ANN is appropriate for the estimations of the similar type of the another set of data. It must be tested the generalization ability of the ANN by using the second set of the data which is never seen by the constructed ANN in the training stage. If the generated outputs in the test stage by using final weights are still close to the desired outputs, it can be confidently concluded that the ANN is convenient of solving this type of problem. For further details of ANN, the authors refer to read the reference (Haykin, 1999).

## 3. Results and Discussion

In this study, after several trials, the number of hidden layer neurons was determined and used as 5 which gives the best results for the problem. All data were divided into two separate parts for training (80%) and test (20%) stages. For the training of the ANN, a back-propagation algorithm with Levenberg–Marquardt (Levenberg, 1944; Marquardt, 1963) was used. By appropriate modifications, final weights are obtained after an acceptable error level between estimated and desired outputs is attained. The inputs of the ANN are the beam energy, atomic number, neutron number and mass number of the beam, target and compound nuclei. The ANN output belongs to the fusion-evaporation reaction cross section (XS). Five hidden neurons were used in one hidden layer. The number of total adjustable weights is 55. The used topology of the ANN is given in Fig.1.

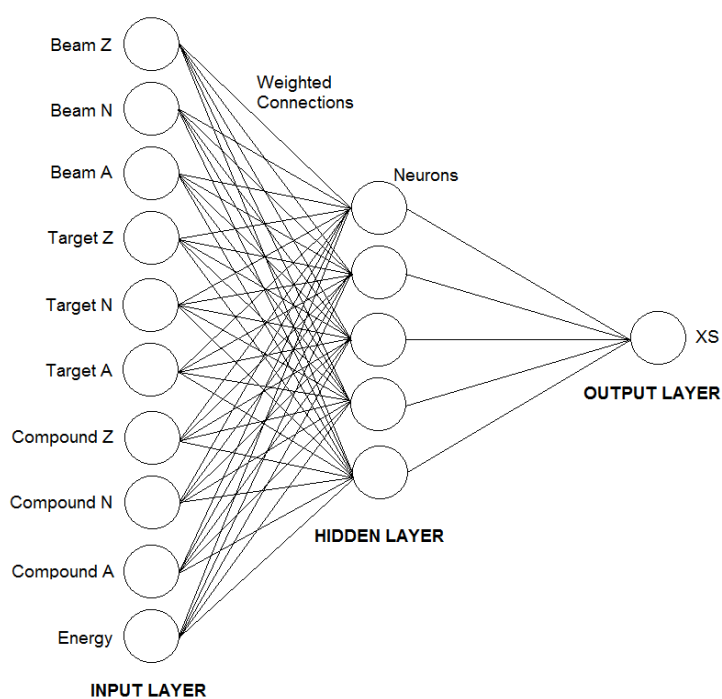

**Fig.1.** The used 10-5-1 type ANN structure for the estimation of fusion-evaporation reaction cross section

After training of the network, the neural network outputs were generated both for training and test data values by using final weights. The average value of the cross sections in the training data is 1029.8. The MSE value for the training data is 18.5 mb which corresponds to about 1.8% deviations from the experimental data values. The minimum and maximum absolute errors are 0.03 and 63.6 mb for the training data. The correlation coefficient is obtained 0.99 in this stage. As is clear also in Fig.2 that ANN method produced very close data points to the highly non-linear experimental cross section values for the fusion-evaporation reactions.

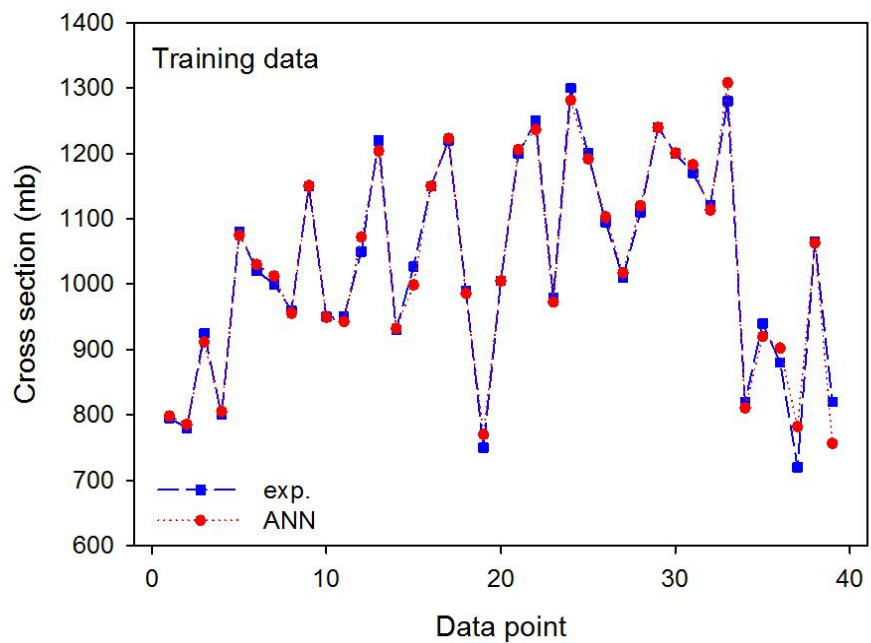

**Fig. 2.** Comparison of experimental cross sections and ANN predictions on the training data for some fusion-evaporation reaction.

The test stage is quite important for seeing the success of the method. By using final weights of the ANN, we have predicted the cross sections for the fusion-evaporation reactions in the test data. The average value of the cross sections in the test data set is 1048.4. The MSE value for the test data is 110.4mb which corresponds to about 10.5% deviations from the experimental data values. The minimum and maximum absolute errors are 4.1 and 187.6mb for the data. The correlation coefficient is obtained 0.72 in this stage which is still high. We have shown the differences for the cross sections between ANN predictions and experimental values in Fig.3. In same figure we have also given similar deviations between most common theoretical cross section PACE results and experimental data. It is clear in figure that the

ANN deviations are lower than the PACE. The MSE value for the PACE results is 122.8mb which are worse than the ANN predictions. The minimum and maximum absolute errors are 6.0 and 237.0mb for the PACE.

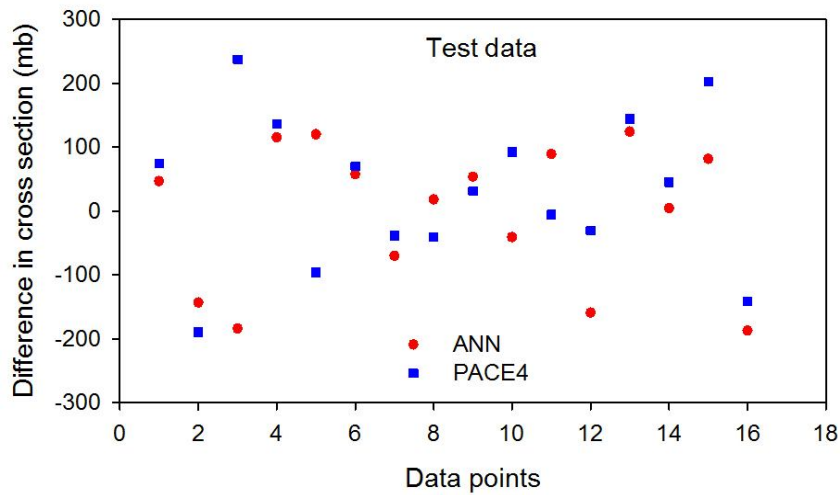

**Fig.3.** Differences of ANN predictions (circle)and PACE4 calculations (square)from experimental cross section values on the test data.

We have given the detail of the reactions in the test data set and corresponding cross section values from different methods such as experiments, ANN predictions and PACE calculations. For 10 of total 15 different fusion-evaporation reactions, the ANN predictions are closer to the experimental values than PACE results (these are marked bold in the table). Whereas for only 5 reactions, PACE give better results.

**Table 1** The experimental, ANN predicted and PACE calculated cross sections for the different reactions with different beam energies used in test stage.

| Beam | | | Target | | | Compound | | | Energy (MeV) | Cross section(mb) | | |
|---|---|---|---|---|---|---|---|---|---|---|---|---|
| Z | N | A | Z | N | A | Z | N | A | | Exp. | ANN | PACE |
| 3 | 4 | 7 | 6 | 6 | 12 | 9 | 10 | 19 | 16.3 | 975 | **928.48** | 901 |
| 4 | 5 | 9 | 14 | 14 | 28 | 18 | 19 | 37 | 23.8 | 950 | **1093.56** | 1140 |
| 5 | 5 | 10 | 7 | 7 | 14 | 12 | 12 | 24 | 35.2 | 1045 | **1229.09** | 808 |
| 6 | 7 | 13 | 6 | 6 | 12 | 12 | 13 | 25 | 18 | 960 | 840.48 | **1056** |
| 6 | 6 | 12 | 12 | 12 | 24 | 18 | 18 | 36 | 33.7 | 1200 | **1085.32** | 1064 |
| 6 | 6 | 12 | 12 | 14 | 26 | 18 | 20 | 38 | 37 | 1300 | **1218.63** | 1098 |
| 6 | 6 | 12 | 14 | 14 | 28 | 20 | 20 | 40 | 34.5 | 960 | 1147.62 | **1102** |
| 7 | 8 | 15 | 13 | 14 | 27 | 20 | 22 | 42 | 45 | 1200 | **1142.80** | 1130 |
| 8 | 8 | 16 | 6 | 6 | 12 | 14 | 14 | 28 | 20.8 | 1060 | **1055.91** | 1015 |
| 8 | 8 | 16 | 14 | 14 | 28 | 22 | 22 | 44 | 38.8 | 1050 | **1032.30** | 1091 |
| 8 | 8 | 16 | 14 | 15 | 29 | 22 | 23 | 45 | 39 | 1260 | **1136.27** | 1116 |
| 9 | 10 | 19 | 14 | 16 | 30 | 23 | 26 | 49 | 44.5 | 1235 | 1181.68 | **1204** |
| 12 | 12 | 24 | 12 | 12 | 24 | 24 | 24 | 48 | 39.3 | 1050 | **1091.32** | 958 |
| 14 | 14 | 28 | 14 | 16 | 30 | 28 | 30 | 58 | 42.8 | 820 | 731.23 | **826** |
| 14 | 14 | 28 | 12 | 12 | 24 | 26 | 26 | 52 | 38 | 760 | 919.29 | **791** |

## 4. Conclusions

Due to the fact that there is limited number of experimental data available in the literature, we have performed ANN method on experimental fusion-evaporation reaction cross section data. The theoretical estimations from different codes should be divided by a factor about 5-10 in order to obtain experimental cross sections. It has been seen that the ANN method is a practical method for the cross section estimations. By using the 10-5-1 ANN topology, the cross sections have been obtained with 1.8% and 10.5% errors on training and test data sets, respectively. The correlation coefficients for training and test data is higher than 0.7. One can confidently use ANN method for the estimation of fusion-evaporation reaction cross section.